\documentclass[12pt,preprint]{aastex}

\shortauthors{T. Hosokawa}
\shorttitle{Do Lyman-Break Galaxies Have Supermassive BHs?}

\begin{document}

\title{Cosmological Growth History of Supermassive Black Holes
and Demographics in the High-{\it z} Universe  \\
: Do Lyman-Break Galaxies Have Supermassive Black Holes?}

\author{Takashi Hosokawa}
\affil{Yukawa Institute for Theoretical Physics, Kyoto University,
Kitashirakawa, Sakyo-ku, Kyoto 606-8502}
\email{hosokawa@yukawa.kyoto-u.ac.jp}

\begin{abstract}
We study the demographics of supermassive black holes (SMBHs) in the local
and high-{\it z} universe. 
We use the continuity equation of the population of SMBHs
as the leading principle. 
We consider three physical processes for the growth of SMBHs:
mass accretion, mergers, and direct formation of SMBHs.
The mass accretion history of SMBHs is estimated by the
 hard X-ray luminosity functions (HXLFs) of AGNs.
First, we compare the mass accretion history at $z>0$ with optical
luminosity functions (OLFs) of QSOs previously studied and that with
HXLFs. We conclude that the constraints on parameters of mass accretion
(energy conversion efficiency, $\epsilon$ and Eddington ratio, $f_{\rm Edd}$)
based on the continuity equation appear to be adequate using
 HXLFs rather than OLFs. The sub-Eddington case ($f_{\rm Edd} < 1$) is
allowed only when we use HXLFs. 
Next, we extend the formulation and we can obtain the upper limit of
the cumulative mass density of SMBHs at any redshifts.
 For an application, we examine if 
Lyman-Break galaxies (LBGs) at $z \sim 3$ already have SMBHs 
in their centers which is suggested by recent observations. 
We tentatively assume the presence of SMBHs in LBGs and that 
their mass, $M_{\rm BH}$ is 
proportional to the stellar mass of LBGs, $M_*$ with the 
mass ratio $\xi = M_{\rm BH}/M_*$. 
If most of LBGs already has massive SMBHs at $z \sim 3$,
the resultant mass density of SMBHs at $z \sim 0$ should exceed the
 observational estimate because such SMBHs should further grow by accretion. 
Therefore, we can set the upper limit of the
probability that one LBG has a SMBHs. 
Since the merger rates and direct formation rates of SMBHs are 
uncertain, we consider two limiting cases : (i)
mergers and/or direct formations are not negligible compared with
mass accretion and (ii) mass accretion is the dominant process to
grow the SMBHs. 
The special conditions should be met in order that a large part of LBGs
have SMBHs in both cases.
In case (i), we may assume the constant parameters of mass 
accretion of AGNs for simplicity. 
Then, large energy conversion efficiency and frequent
mergers and/or direct formations at $z>3$ are needed so that
a large part of LBGs have SMBHs with $\xi = 0.002$ to 0.005.
Whereas, in case (ii), energy conversion efficiency should be mass
dependent and the constraint is strict;
the fraction of LBGs which have SMBHs must be less than 10-40\%. 
In both cases, the possibility that nearly all LBGs have SMBHs with 
large mass ratio, such as $\xi \geq 0.005$, is reliably ruled out.
\end{abstract}

\keywords{black hole physics --- galaxies: active --- galaxies :
evolution --- galaxies: luminosity function, mass function --- galaxies
: nuclei --- quasars: general}

\section{Introduction}

Today, there is a wide consensus that almost all local galaxies harbor
supermassive black holes (SMBHs) in their centers (e.g. 
Begelman 2003, Kormendy \& Richstone 1995).  We can estimate the
mass of these SMBHs with dynamical motion of gases or stars around
the center of galaxies (e.g. Ghez et al. 1998, Richsotne et al. 1998, 
Gebhardt et al.2003). 
It is well-known that there are some correlations between central SMBHs 
and its galactic bulges. 
For example, the mass of SMBHs is proportional to the mass
of their galactic bulges;  
$M_{\rm BH}/M_{\rm bulge} \sim 0.001-0.006$, 
though its tightness and linearity is still under discussion 
(Kormendy \& Richstone 1995, Maggorian et al. 1998, Laor 2001, 
McLure \& Dunlop 2002, Marconi \& Hunt 2003).
Another tight correlation is so-called $M_{\rm BH}-\sigma$
relation, where $\sigma$ is the stellar velocity dispersion in the bulges.
This relation is expressed as the power-law form, $M_{\rm BH} \propto
\sigma^{3.8-4.8}$ (e.g. Gebhardt et al. 2000, Ferrarese \& Merritt 2000,
Tremaine et al. 2002). 
Recently, Ferrarese (2002) and Baes et al. (2003)
argue that SMBH mass is related to the dark halo mass beyond the bulge, 
$M_{\rm BH} \propto M_{\rm DM}^{1.27-1.82}$.
Graham et al. (2001) point that there is another correlation
that bulges with more massive SMBHs have steeper central cusps.
In spite of discoveries of many correlations and great efforts which
have been devoted to clarify the physical meanings of these correlations
 (e.g. Silk \& Rees 1998, Ostriker 2000, Umemura 2001, 
Adams, Graff \& Richstone 2001),
we do not have a consensus yet regarding physical link
between the formation process of SMBHs and that of
their host galaxies (bulges).

To understand how the central SMBHs formed and how their formation
history is related to that of galaxies,
we must at least know when SMBHs were formed in galaxies.  
Therefore, it is very important to know whether high-{\it z} (not active) galaxies 
already had SMBHs or not.     
In this respect, it is interesting to note that
the so-called Lyman-Break technique, for example, has made it possible
to discover a significant number of high-{\it z} ($> 3$) galaxies,
i.e., Lyman-Break Galaxies (LBGs, Steidel et al. 1996). 
These LBGs are widely known as high-{\it z} starburst galaxies and 
their typical starformation rate is $10^2 - 10^3 M_\odot {\rm yr}^{-1}$.
Papovich et al. (2001) and Shapley et al. (2001)
applied the population synthesis models to infer UV-to-optical
spectrum of LBGs at their rest frame and examined various properties
of LBGs. Despite uncertainties their results are consistent
at some points; 
LBGs are typically young (their ages are several hundred Myr since last
star-formation events) and small (their typical stellar mass is $\sim 10^{10} M$),
though there were some old LBGs.   
 
Ridgway et al. (2001) point that 
the sizes and magnitudes of the host galaxies of the radio-quiet QSOs 
at $z \sim 2-3$ are similar to those of LBGs in HST imaging study.
Steidel et al.(2002) observed about 1000 LBGs 
and argue that 3\% of these LBGs are optically faint AGNs based on their UV
spectrum. Nandra et al.(2002) confirmed by using {\it Chandra} observatory
that the same number fraction, 3\%, of LBGs are extremely bright in the 
rest-frame hard X-ray band, and asserted that these LBGs show AGN activity.
They discussed that this number fraction, 3\% may reflect the duty
cycle of mass accretion to SMBHs in LBGs. 
That is, if the typical age of LBGs is 300 Myr (e.g. Shapley et
al. 2001) and the accretion timescale is $10^7$ yr,
remaining 97\% of LBGs may have inactive SMBHs. 
This situation may be compared with the local universe.
In the local universe, almost all galaxies harbor SMBHs in their
centers, but there are much inactive galaxies than AGNs.
The pioneering work (Dressler, Thompson \& Shectman 1985) found that 
the number franction of AGNs to all galaxies is only several \%, 
though this franction recently increases to 20-40 \% due to the 
accurate observations (Kauffmann et al. 2003, Miller et al. 2003).
Furthermore, Vestergaard (2003) and Vestergaard (2004)
discusses that the LBGs with AGN activity
may have SMBHs of order $10^8 M_\odot$ from the virial mass estimate using
C IV line width. If these LBGs have typical stellar mass of $M_* \sim 10^{10}
M_\odot$, the mass ratio becomes $M_{\rm BH}/M_* \sim 0.01$, which is larger
than $M_{\rm BH}/M_{\rm bulge} \sim 0.001-0.006$ of local galaxies. 
Incidentally, normal galaxies and AGNs have the same mass ratio,
 $M_{\rm BH}/M_{\rm bulge}$, in the local universe (Wandel 2002).
Whereas, Granato et al. (2001) found that LBGs may be equivalent 
to the galaxies which are in the pre-AGN phase based on their 
semi-analytic model of the joint formation of QSOs and galaxies.
Kawakatu et al.(2003) argue that it may be physically difficult to feed 
a black hole in optically thin galaxies, such as LBGs, based on 
the idea by Umemura (2001). 
In the local universe, almost all the SMBHs observationally accompany 
the stellar spheroidal (bulge) component (M33, which is the bulgeless galaxy,
appears to have no SMBHs (Gebhardt et al. 2001)), therefore, 
the bulge component may be indispensable to form a SMBH.
If LBGs are still in the initial phase of forming the bulge component
(Matteucci \& Pipino 2002), LBGs may not have SMBHs.

Of course, it is difficult to resolve the central parsec scale region of LBGs
and search the (inactive) SMBHs observationally.
In this paper, we theoretically examine the possibility that LBGs have
SMBHs from the viewpoints of the demographics. 
We consider the cosmological accretion history 
derived from the observed hard X-ray (2-10keV) luminosity functions
(HXLFs) of AGNs at $z>0$ (Ueda et al. 2003, hereafter UAOM03).
Yu \& Tremaine (2002) (hereafter YT02) use the continuity equation
for the number density of SMBHs (see also Soltan 1982, Small \& Blandford 1992) 
and show that the accreted mass estimated with optical LFs (OLFs) of
QSOs at $z>0$ is comparable with the SMBH mass density of the local universe.
However, we can see only a part of various populations of AGNs in the
optical band, thus the analysis with hard X-ray LFs, which include
absorbed AGNs, is important.
We consider the cosmological accretion history estimated with HXLFs and compare
it with the accretion history estimated with OLFs (YT02).
Furthermore, we basically extend the formulation of YT02 and crudely estimate
how many and how massive SMBHs are allowed to exist in the high-{\it z} universe.
If most LBGs already had very massive SMBHs at $z \sim 3$, 
the accreted mass into SMBHs until $z \sim 0$ should exceed
the BH mass density of the local universe. 
In this way, we will be able to set constraints on the mass density of
SMBHs in LBGs.

Below, we use the continuity equation of SMBH population as the leading
principle. In \S2, we first transform the continuity equation to the useful
integrated form (\S2.1). 
We then compare the cosmological accretion
history at $z>0$ with the local population of SMBHs and
constrain the physical parameters; 
the energy conversion efficiency, $\epsilon$ and the Eddington ratio, 
$f_{\rm LBG} (\equiv L_{\rm bol}/L_{\rm Edd}$) (\S2.2). 
In this time, we compare the accretion history with HXLFs and that with
OLFs. 
In \S3, we extend YT02's formulation and derive the upper limit
of the population of SMBHs at $z \sim 3$ (\S3.1). Furthermore, we
consider the tentative SMBHs in LBGs with mass ratio,
$\xi \equiv M_{\rm BH}/M_*$ ($M_*$ is the stellar mass of LBGs),
and crudely estimate the upper limit of the probability
that one LBG has a SMBH (\S3.2). 
In \S4, we summarize and discuss the results. 
Throughout this paper, we adopt the $\Lambda$-dominant universe
and the cosmological parameters are $(\Omega_M,\Omega_\Lambda,h)=
(0.3,0.7,0.7)$.

\section{Cosmological Accretion History v.s. Local Population of 
Supermassive Black Holes}

\subsection{Continuity Equation of the Population of SMBHs}

We begin with the continuity equation for the number density of SMBHs:
$n(M,t) \equiv dN/dM(M,t)$; that is
\begin{equation}
\frac{\partial n}{\partial t}+
\frac{\partial [n \cdot \langle \dot M \rangle ]}{\partial M}
= \gamma_{\rm merge}(M,t)+\gamma_{\rm form}(M,t)
\label{conti}
\end{equation}
(Soltan 1982, Small \& Blandford 1992, YT02). 
Hereafter, we simply express the black
hole mass as $M$. The source terms on the right-hand side, 
$\gamma_{\rm merge}(M,t)$ and $\gamma_{\rm form}(M,t)$, represent 
the SMBH-SMBH merger rates (e.g. Haehnelt 1994)
and the rates of the direct formation 
of SMBHs by, say, collapse of the supermassive stars (SMS) 
($\sim 10^5 M_\odot$, or may be more massive). 
We can express these terms explicitly. The merger term is,
\begin{eqnarray} 
\lefteqn{ \gamma_{\rm merge}(M,t) = } \nonumber \\ 
&&
\frac12
\int_0^\infty\!\!\!\!\!\!dM_1 \int_0^\infty\!\!\!\!\!\!dM_2 \ \
 \Omega(M_1,M_2) \
\left[ \delta(M-M_1-M_2)-\delta(M-M_1)-\delta(M-M_2) \right],
\label{mg}
\end{eqnarray}
where $\Omega(M_1,M_2)$ is the distribution function which represents
the merger rates among the SMBHs of mass $M_1$ and $M_2$.
The factor 1/2 is necessary to correct the double counts of the merger rates
in the integral in eq.(\ref{mg}).
Several recent studies investigate the distribution function of
mergers, $\Omega(M_1,M_2)$ with the semi-analytic model of
galaxies and QSOs (e.g. Kauffmann \& Haehnelt 2000, Volonteri, Haardt
\& Madau 2003). However, the key problem of the SMBH merger is the
timescale. That is, the timescale to refill the "loss cone" of the SMBH
binary (see Begelman et al. 1980 or Milosavijevi\'c \& Merritt 2003). 
Although the hardening timescale of the binary have been actively
studied, this problem is still under discussion (Makino 1997, 
Milosavijevi\'c \& Merritt 2001, Milosavijevi\'c \& Merritt 2003, 
Makino \& Funato 2004). Therefore, we treat $\Omega(M_1,M_2)$ as the 
unknown function.
Here, we assume that the total mass of SMBHs is conserved through 
merger events.
As for $\gamma_{\rm form}$, we only note that this numerical value is
not negative,
\begin{equation}
\gamma_{\rm form}(M,t) \geq 0.
\end{equation}
Recently, the gravitational collapse of SMS is being studied with 
the general relativistic numerical simulations (e.g. Shapiro \&
Shibata 2002, Shibata \& Shapiro 2002). However, it is still
uncertain how and when the initial condition of the simulations is realized.
Eventually, these source terms are highly uncertain and we do not know
whether these terms exist or not 
(the merger term will be at least non-negligible (e.g. Merritt \& Ekers 2003)).  
YT02 successfully avoid this difficulty by multiplying eq.(\ref{conti}) 
by the function, 
\begin{equation}
f(M,M') = \left\{
          \begin{array}{rl}
          0,& \quad \mbox{for $M'<M$} \\
          M'-M,& \quad \mbox{for $M'>M$} 
          \end{array}\right. 
\label{f}
\end{equation}
and integrate by $M'$ from 0 to $\infty$ and by $t$ from $0$ to
$t_0$ (cosmological time at $z=0$), assuming that
$n(M,t) \to 0$ in the limit of $t \to 0$ and $M' \to \infty$.
The result is
\begin{equation}
G_{\rm local}(M,t_0) = G_{\rm acc}(M,t_0)
+G_{\rm merge + form}(M,t_0).
\label{geq}
\end{equation}
Here, $G_{\rm local}(M,t_0)$, $G_{\rm acc}(M,t_0)$ 
and $G_{\rm merge + form}(M,t_0)$ are defined by
\begin{equation}
G_{\rm local}(M,t_0) \equiv \int_M^\infty (M'-M) n(M',t_0) \ dM',
\end{equation}
\begin{equation} 
G_{\rm acc}(M,t_0) \equiv \int_M^\infty \ dM' \int_0^{t_0}
n(M',t)  \langle \dot M' \rangle \ dt,
\end{equation}
\begin{equation}
G_{\rm merge + form}(M,t_0) \equiv
\int_M^\infty \int_0^{t_0}  
(M'-M) \left( 
\gamma_{\rm merge}(M',t) + \gamma_{\rm form}(M',t)
\right)
\ dM' dt,
\end{equation}
where $\langle \dot M' \rangle$ is the average mass accretion rate
into the SMBHs of mass $M'$.
Since $G_{\rm merge + form}(M,t_0)$ is 0 or positive, 
we obtain the inequality,
\begin{equation}
G_{\rm local}(M,t_0) \geq G_{\rm acc}(M,t_0).
\label{ineq}
\end{equation}

These mathematical derivation of eq.(\ref{ineq}) is the same as
that in YT02, whereas we explain the physical meaning of eq.(\ref{ineq}) here.
To do so, we transform inequality (\ref{ineq}) as
\begin{equation}
\int_M^\infty M' n(M',t_0) \ dM' \geq 
\int_M^\infty \ dM' \int_0^{t_0} n(M',t) \langle \dot M' \rangle dt
+ M \int_M^\infty n(M',t_0) \ dM'.
\label{mod_ineq}
\end{equation}
In this form, we can understand the meanings of each term.
Clearly, the left-hand side of eq.(\ref{mod_ineq}) represents
the mass density contained by SMBHs more massive
than $M$ at $z=0$. 
The first term of the right-hand side
represents the mass accreted into the SMBHs more massive 
than $M$ per unit volume during $0 \leq t \leq t_0$ (i.e. $z \geq 0$).
The second term on the right-hand side represents
another contribution; 
that is, the mass density of SMBHs at the moment that SMBHs become 
more massive than $M$. 
In our consideration, the SMBHs more massive than $M$ can form
through the mass accretion, mergers, or gravitational collapse of the SMS.
The mass of a SMBH continuously grows by the mass accretion, 
whereas the mass of a SMBH
grows discontinuously by the mergers and/or direct formation.
Therefore, the mass of SMBHs at the moment that SMBHs become more massive
than $M$ is equal to $M$ for the case with the mass accretion, 
whereas it is more than $M$ otherwise.
The total number of SMBHs which became more massive than $M$ until $z=0$
is necessarily larger than $\int_M^\infty n(M',t_0) dM'$ since the
mergers reduce the number of SMBHs.
Totally, the lower limit of the contribution of the mass density of 
SMBHs at the moment that SMBHs become more massive than $M$ is
$M \int_M^\infty n(M',t_0) dM'$, which is the second term on the
right-hand side of eq.(\ref{mod_ineq}).
The physical meaning of this term may be considered as
the ``advection'' of SMBH mass across the mass $M$
in terms of the continuity equation of the population of SMBHs.
This ``advection'' term is the heart of this inequality.

We can calculate $G_{\rm local}(M,t_0)$ with the local mass function
of SMBHs. We adopt the mass function of SMBHs given by Aller \& 
Richstone (2002) (hereafter AR02) based on the luminosity functions 
of galaxies at $z=0$ for each morphology. 
They convert the local LFs of normal galaxies to the mass function
of SMBHs with some empirical relations among total luminosity,
bulge luminosity, velocity dispersion and SMBH mass.
One of these relations is $M-\sigma$ relation mentioned above.
AR02 derive the two local mass functions of SMBHs for
different galactic LFs ; mass function (a) is derived from LFs 
by Marzke et al. (1994), whereas mass function (b) is derived from LFs 
by Madgwick et al. (2002). 
Below, we show the results of the calculation for the cases with mass function (a)
and refer the results of the other cases in text.
We use the fitting formulae for the mass functions of SMBHs given by AR02. 
With the mass functions (a) and (b), the local mass density of SMBHs more 
massive than $10^6 M_\odot$ is 
$(4.8 \pm 1.6)h^2 \times 10^5 M_\odot {\rm Mpc}^{-3}$ and
$(6.9 \pm 1.4)h^2 \times 10^5 M_\odot {\rm Mpc}^{-3}$ respectively.
YT02 adopt the LFs based on the SDSS (Sloan Digital Sky Survey) and
their calculated value is $(2.5 \pm 0.4) \times 10^5 
M_\odot {\rm Mpc}^{-3}$ with $h=0.65$. 
Salucci et al. (1999) and Merritt \& Ferrarese (2001) also
obtain similar results.

To calculate $G_{\rm acc}(M,t_0)$, there is a
difficulty in that we do not know the number density of SMBHs at
cosmological time $t$, $n(M,t)$, and the average accretion rate
of SMBHs, $\langle \dot M \rangle$.
Here, we assume that the mass accretion into SMBHs occurs mainly at
bright AGN phase.
The average accretion rate of AGNs
is assumed to be 
\begin{equation}
\langle \dot M \rangle = \frac{(1-\epsilon)L_{\rm bol}}{\epsilon c^2},
\label{mdot}
\end{equation}
where $L_{\rm bol}$ is the bolometric luminosity of AGNs and
$\epsilon$ is the energy conversion efficiency.
With eq.(\ref{mdot}), we can express $G_{\rm acc}(M,t_0)$ as
\begin{equation}
G_{\rm acc}(M,t_0)=
\int_{L(M)}^\infty dL' \int_0^{t_0} \frac{d\Phi(L',t)}{dL'}
\frac{(1-\epsilon)L_{\rm bol}}{\epsilon c^2}\ dt.
\label{gacc}
\end{equation}
Here, $d\Phi(L,t)/dL$ is the LFs of AGNs and we use observed HXLFs 
(UAOM03). Very recently, UAOM03 build the HXLFs of 
AGNs over the wide luminosity range, 
$\log (L_x (2-10 {\rm keV}) \ {\rm erg/s}) = 41.5-46.5$, and wide redshift
range, $z=0-3$, from the combination of surveys with {\it HEAO1}, 
{\it ASCA} and {\it Chandra} satellites.
The shape of HXLFs of AGNs at $z=0$ are fitted by so-called double power-law
function well. That is,
\begin{equation}  
\frac{d \Phi(L_x,z=0)}{d \log L_x}
= A \left[ 
    (L_{\rm X}/L_*)^{\gamma 1} + (L_{\rm X}/L_*)^{\gamma 2}
\right]^{-1} .
\end{equation}
In the above expression, $\gamma 1$ and $\gamma 2$ represent the inclinations
of LFs at the lower and higher luminosity ends respectively, and 
$L_*$ represent the luminosity of the ``knee'' of LFs. 
The redshift evolution of HXLFs are described as the
luminosity-dependent density evolution (LDDE), which is suggested by
Miyaji et al. (2001) for the evolution of the soft X-ray luminosity 
functions of AGNs. The LDDE model is expressed as
\begin{equation}
\frac{d \Phi(L_{\rm X},z)}{d \log L_{\rm X}} 
= \frac{d \Phi(L_{\rm X},0)}{d \log L_{\rm X}}
\times e(z, L_{\rm X}),
\end{equation}
where
\begin{equation}
e(z,L_x) = \left\{
          \begin{array}{lr}
          (1+z)^{p1},& \quad \mbox{for $z < z_c (L_{\rm X})$} \\
          e(z_c)[(1+z)/(1+z_c(L_{\rm X}))]^{p2},& 
                            \quad \mbox{for $z \geq z_c (L_{\rm X})$}, 
          \end{array}\right. 
\end{equation}
and
\begin{equation}
z_c(L_{\rm X})= \left\{
          \begin{array}{lr}
          z_c^*,& \quad \mbox{for $L_{\rm X} \geq L_a$} \\
          z_c^* (L_{\rm X}/L_a)^\alpha,& \quad \mbox{for $L_{\rm X} < L_a$}. 
          \end{array}\right. 
\end{equation}
The parameters which represent redshift dependence of
$e(z,L_{\rm X})$ are $p1 > 0$ and $p2 < 0$. Then, $z_c(L_{\rm X})$ 
 corresponds to the peak redshift at which the number density of AGNs with 
luminosity of $L_{\rm X}$ is the maximum. 
The cut-off redshift, $z_c(L_{\rm X})$, increases as the luminosity increases
($\alpha > 0$) (see Fig.10 of UAOM03). 
That is, the number density of the more luminous AGNs
reaches the maximum at the higher redshift and decrease as $z
\rightarrow 0$ (see Fig.12 of UAOM03). 
We simply extrapolate the fitting formulae given by above LDDE model
at $z>3.0$.

To calculate $G_{\rm acc}(M,t_0)$, we need the relation between
the hard X-ray luminosity, $L_{\rm X}$ and the bolometric luminosity, $L_{\rm
bol}$. We assume the relation as $L_{\rm bol} = C_{\rm X} L_{\rm X}$, where
$C_{\rm X}$ is constant value $C_{\rm X} = 25.0$ (Alonso-Herrero et
al. 2002). Generally, the ratio of B-band (4400 \AA), hard X-ray (2-10 keV) 
luminosity to the bolometric luminosity, $C_{\rm B} = L_{\rm B}/L_{\rm
bol}$ and $C_{\rm X}$ should depend on the mass accretion rate and the 
black hole mass. This may be important to compare the local mass
function of SMBHs with the mass accretion history estimated with OLFs
and that estimated with HXLFs. We will discuss this issue in the next subsection
in some details.
We suppose that the QSOs radiate as $L_{\rm bol}=f_{\rm Edd} L_{\rm
Edd}=1.5 \times 10^{38} f_{\rm Edd} (M/M_\odot) (\mbox{erg/s})$. 
The parameters here are the energy conversion efficiency, $\epsilon$, 
and the Eddington ratio, $f_{\rm Edd}$.

\subsection{Comparison between the Mass Accretion History with HXLFs and
  that with OLFs}
 
Here, we compare the mass accretion history estimated with HXLFs 
and that with OLFs based on the continuity equation. 
For both cosmological accretion history with HXLFs and that with OLFs, 
the relation to the local mass function of SMBHs can be investigated
with ineq.(\ref{ineq}) derived from the continuity equation. 

YT02 noted that their calculated $G_{\rm acc}(M,t_0)$ with OLFs of QSOs
with the typical parameter, $\epsilon = 0.1$ and $f_{\rm Edd}=1.0$
is larger than $G_{\rm local}(M,t_0)$, especially in the large mass
range, $M > 10^7 M_\odot$ (inequality (\ref{ineq}) is not satisfied). 
That is, there is the problem of the overaccretion compared to the local
mass function of SMBHs.
YT02 discussed the possible origin of this discrepancy. 
For example, they argue that the optically bright QSOs may radiate at
large mass-to-energy conversion efficiency, $\epsilon$ (e.g. $\epsilon
\geq 0.2$). However, we can see only a part of the light emitted in the 
mass accretion process to SMBHs in the optical band.
This is partly because the optical luminosity emitted through the 
mass accretion can be absorbed by obscuring torus (Type II AGNs) 
(e.g. Antonucci 1993).
We can miss the reddened QSOs due to the dust extinction in the 
host galaxies when the typical color selection criteria (e.g. UV excess) 
is adopted to build the QSO LFs (e.g. Webster et al. 1995). 
Thus, we calculate $G_{\rm acc}(M,t_0)$ with HXLFs and compare it
to $G_{\rm acc}(M,t_0)$ with OLFs and to $G_{\rm local}(M,t_0)$. 
Of course, as YT02 noted, it seems to make the problem worse
to use the HXLFs. This is because HXLFs include extra populations 
of AGNs besides the optically bright AGNs. However, this is not the
case. The redshift evolution of HXLFs and that of OLFs are different,
then we can make the situation better (see below).

The upper panel of Fig.1 represents the results with the local mass function
of SMBHs (a).  
For OLFs of QSOs, we adopt ones given by Boyle et al.(2000)
at $0.35 < z < 2.3$ derived from about 6,000 QSOs detected by 2QZ survey
and simply extrapolate it at $z >2.3$.
We assume the constant bolometric correction, $C_{\rm B} = 11.8$ (Elvis
et al. 1994). YT02 use the same OLFs and adopt the same assumptions. 
Like YT02's result, our results show that the inequality (\ref{ineq}) 
is not satisfied in all mass range (especially at high mass end) with
OLFs and typical parameters, $(\epsilon, f_{\rm Edd}) = (0.1,1.0)$.
If we use mass function (b), the calculated $G_{\rm local}(M,t_0)$ 
is larger than (a) by a factor 1.4, but equation (\ref{ineq}) is not still 
satisfied at $M_{\rm BH} \geq 10^{7.5} M_\odot$. 

However, Fig.1 also shows that calculated $G_{\rm acc}(M,t_0)$ with
HXLFs with the same parameter set do not exceed $G_{\rm local}(M,t_0)$ 
in $M < 10^{8.5} M_\odot$.
At first sight, this looks strange.
Since in HXLFs optically absorbed AGNs and reddened AGNs are considered
besides the optically bright QSOs, $G_{\rm acc}(M,t_0)$ 
calculated based on HXLFs should become 
larger than $G_{\rm acc}(M,t_0)$ based on OLFs. 
The origin of this puzzle lies in our assumption on $C_{\rm B}$ and
$C_{\rm X}$. 
We assume that both of $C_{\rm B}$ and $C_{\rm X}$ are constant;
$C_{\rm B}=11.8$ and $C_{\rm X} = 25.0$. 
If this were actually the case, the ratio
between B-band luminosity and hard X-ray luminosity, $L_{\rm B}/L_{\rm
X}$, must be constant. However, UAOM03 compare their HXLFs of only optical
type-I AGNs to OLFs by Boyle et al. (2000) and find that $L_{\rm X}$ is not
proportional to $L_{\rm B}$. They obtained a power-law relation between
the 2keV and $2500\AA$ luminosities, $l_{\rm X} \propto l_{\rm O}^{0.7}$
in the $(\Omega_M,\Omega_\Lambda,h)=(1.0,0.0,0.5)$ universe.
This relation is approximately transformed to the relation between 
$L_{\rm X} (2-10 {\rm keV})$ and $M_{\rm B}$ as
\begin{equation}
M_{\rm B} = -\frac{7}{2} \log[L_{\rm X} (h/0.5)^{-2} \ ({\rm ergs/s})] + 131.5.
\label{mb_lx}
\end{equation}
That is, $L_{\rm B} \propto L_{\rm X}^{1.4}$. 
Therefore, the assumption that both of $C_{\rm B}$ and $C_{\rm X}$ 
are constant is not justified. 
As noted by UAOM03, if $L_{\rm B}/L_{\rm X}$ were constant, the redshift
evolution of OLFs would be more quick than that of HXLFs.
The bright end of OLFs grows more quickly than HXLFs as $z$
increases.  Then, calculated $G_{\rm acc}(M,t_0)$ based on OLFs is larger
than that based on HXLFs. 
Kawaguchi, Shimura \& Mineghige (2001)
build a disk-corona model for the broad-band spectral energy
distributions (SEDs) of AGNs. Based on their model, Hosokawa et
al. (2001) and Alonso-Herrero et al. (2002) calculated how
$C_{\rm B}$ and $C_{\rm X}$ depend on the black hole mass, $M$ and on
the mass accretion rate, $\dot M$. Though their dependence on $M$ and
$\dot M$ is not simple, $C_{\rm X}$ comparatively remains constant
for varying $M$ and $\dot M$.
Therefore, we adopt constant $C_{\rm X}=25.0$ as noted above and use
this value for the calculations below. 
Incidentally, if $L_{\rm X}$ is proportional to the bolometric
luminosity, $L_{\rm bol}$, eq.(\ref{mb_lx}) may be expressed as 
$C_{\rm B} \propto L_{\rm bol}^{-0.4} \propto M^{-0.4}$
with constant Eddington ratio.
Alonso-Herrero et al. (2002) show the similar dependence,
$C_{\rm B} \propto M^{-0.25}$ based on 
Kawaguchi, Shimura \& Mineshige (2001), though the dependence is 
somewhat weaker. 

Even if we use HXLFs, the calculated $G_{\rm acc}(M,t_0)$ with
$(\epsilon,f_{\rm Edd})=(0.1,1.0)$ becomes larger than 
$G_{\rm local}(M,t_0)$ in the high mass end ($M > 10^{8.5} M_\odot$).
However, this may be an inevitable result.
We calculate $G_{\rm local}(M,t_0)$  with the local mass function of
SMBHs derived with the Schechter-type LFs of galaxies, 
which declines exponentially at the bright end,
whereas we calculate $G_{\rm acc}(M,t_0)$ with the HXLFs
which is power-law function at the bright end. 
YT02 point that the QSO LFs at the bright end or $M-\sigma$ relation
at the high mass end may be uncertain. 
Then, we use the energy conversion efficiency, $\epsilon$, and
Eddington ratio, $f_{\rm LBG}$ as constant free parameters to calculate 
$G_{\rm acc}(M,t_0)$, and search the region of the parameter space
of $(\epsilon,f_{\rm LBG})$ where inequality eq.(\ref{ineq}) is
satisfied only at $M \leq M_0$ for given $M_0$. We choose 
$M_0=10^{8.0} M_\odot$ and $M_0=10^{8.5} M_\odot$. 
Of course, the allowed parameter space
becomes narrower as $M_0$ increases, since the discrepancy we show in Fig.1 
is more prominent at the high mass end.

Fig.2 represents the results for the cases with the local mass function 
of SMBHs (a). 
The allowed region for each $M_0$ is above the labeled lines respectively. 
We set the upper limit of $\epsilon$ as $\epsilon = 0.3$, 
which is the maximum value for the thin-disk accretion models (Thorne
1974). 
If we use mass function (b), the allowed region for each $M_0$
becomes a little wider than (a).
This is consistent with
that the calculated $G_{\rm local}(M,t_0)$ with (b) is larger 
and it is easier to satisfy the inequality (\ref{ineq}) than 
the cases with mass function (a). 
As shown in Fig.2, to satisfy ineq. (\ref{ineq})
at $M \leq 10^{8.5} M_\odot$, the allowed region is not so large
if we demand $f_{\rm Edd} \leq 1$. If $f_{\rm Edd} \sim 1$, $\epsilon$
can be 0.1-0.3. However, only the large energy conversion efficiency
($\epsilon \sim 0.3$) is possible as $f_{\rm Edd}$ decreases 
($f_{\rm Edd} \sim 0.3$).
This conclusion will not change, even if we adopt mass function (b). 
In Fig.\ref{fig.search}, we show the same parameter regions when we use
OLFs of QSOs to calculate $G_{\rm acc}(M,t_0)$. To satisfy 
inequality (\ref{ineq}) to high mass end, high Eddington ratio 
($f_{\rm Edd} > 1$) and high energy conversion efficiency 
($\epsilon > 1$) are needed. As shown above, the cosmological
accretion history estimated with HXLFs naturally satisfy
ineq. (\ref{ineq}) to high mass end with typical parameters 
(especially $f_{\rm Edd} < 1$) and constant luminosity ratio, $C_{\rm
X}=25.0$. Therefore, we conclude that the accretion history with HXLFs is more 
plausible than that with OLFs. Below, we continue to use HXLFs to
estimate the cosmological accretion history of SMBHs.
 
\begin{figure}
\figurenum{1}
\label{fig.gcalc}
\plotone{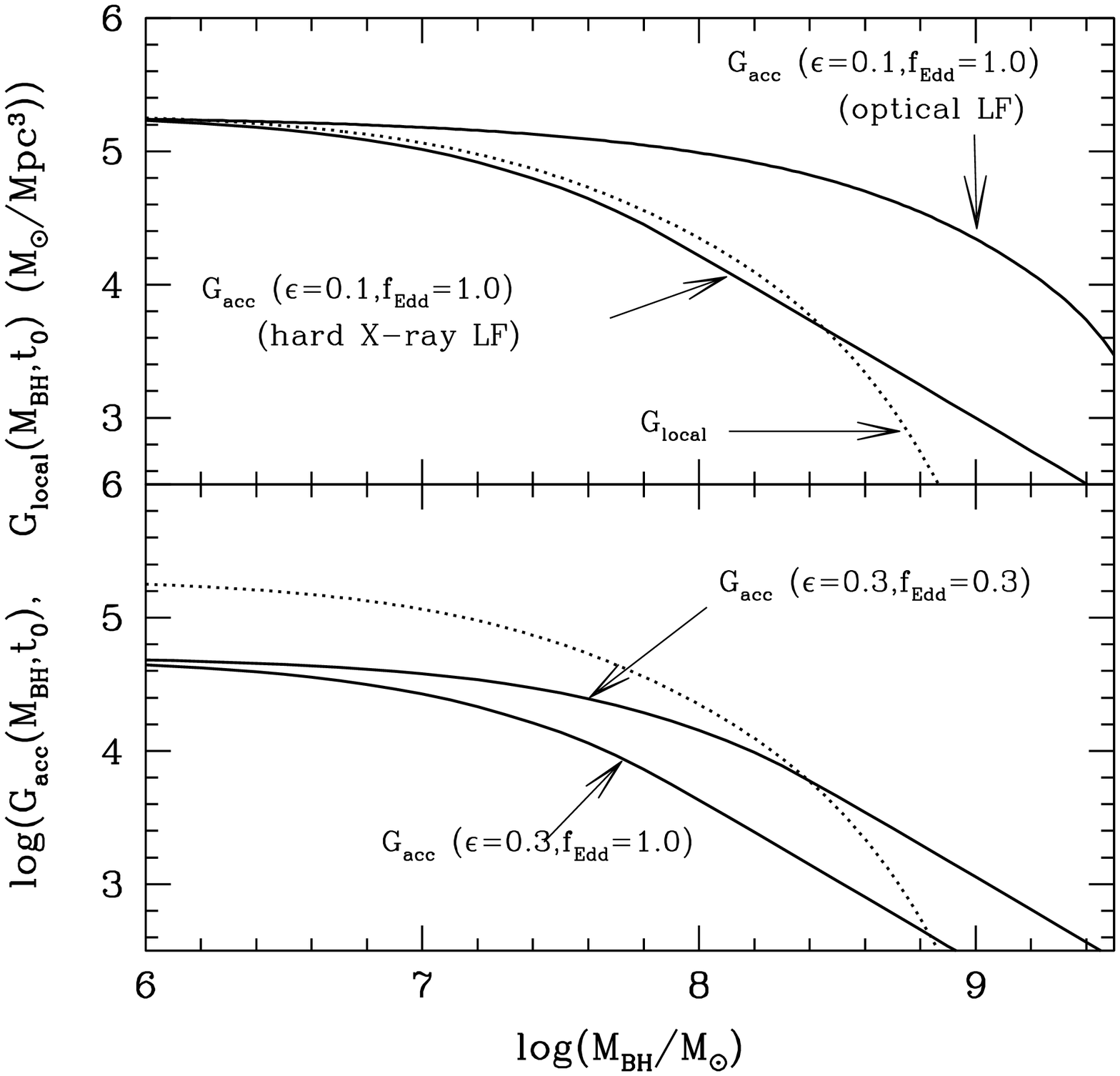}
\figcaption{Calculated $G_{\rm acc}(M,t_0)$ and $G_{\rm local}(M,t_0)$ 
using the local mass function of SMBHs (a) (see text in detail).
{\it Upper panel}: $G_{\rm acc}(M,t_0)$ 
with the typical parameter set, $(\epsilon,f_{\rm Edd})=(0.1,1.0)$ using
OLFs of QSOs and HXLFs of AGNs respectively, 
and $G_{\rm local}(M,t_0)$.
{\it Lower panel}: $G_{\rm acc}(M,t_0)$ with the parameter sets
 indicated by points 2 ($(\epsilon,f_{\rm Edd})=
(0.3,1.0)$), 3 ($(0.3,0.3)$) in Fig.\ref{fig.search}. 
$G_{\rm local}(M,t_0)$ is the same as upper panel.}
\end{figure}

In the lower panel of Fig.1, we plot
the calculated $G_{\rm acc}(M,t_0)$ at points 2 and 3 in Fig.2.
As shown in Fig.\ref{fig.gcalc}, if we adopt constant parameters,
the quantity, $G_{\rm acc}(M,t_0) - G_{\rm local}(M,t_0)$, increases
as $M_{\rm BH}$ decreases to satisfy ineq. (\ref{ineq}) upto high-mass
end (large $M_0$).
From eq.(\ref{geq}), this means that the contribution of mergers and/or
direct formation of SMBHs becomes important at the low mass end. 
If more massive halos possess more massive SMBHs, this may be the case
since mergers between small halos occur
more frequently compared with mergers between large halos under the CDM
cosmology.  

In the next section, we consider the possibility that LBGs have 
inactive SMBHs as in the case of the normal galaxies at $z=0$. 
To do so, we calculate the accreted mass into SMBHs during $0<z<3$.
In this case, we should choose parameter set of $(\epsilon,f_{\rm LBG})$
so that ineq.(\ref{ineq}) is satisfied. Therefore, we use 
$(\epsilon,f_{\rm LBG})$ indicated by points in Fig.2. 
As discussed above, with constant parameters,
contribution of the mergers and/or direct formation of SMBHs to grow 
SMBHs becomes important in the low mass end. Below, we adopt the constant
parameters for the case of non-negligible contribution of mergers
and/or direct formation of SMBHs compared with mass accretion.

\section{Demographics of SMBHs in the High-{\it z} Universe}

\subsection{Upper Limit of the Cumulative Mass Density of SMBHs at
  High-{\it z} Universe}

In this section, we investigate the possibility that LBGs have
SMBHs. 
Our approach is the statistical one. 
That is, if there were already too many or too massive SMBHs in the universe
of $z \sim 3$, the resultant mass density of SMBHs at $z \sim 0$ should
exceed the observed local mass density of SMBHs,
that is a contradiction, because SMBH mass density can never decrease.
The basic formulation is similar to that presented in the previous
section. Here, we multiply the continuity equation, eq.(\ref{conti}),
by $f(M,M')$ defined by eq.(\ref{f}) 
and integrate it
by $M'$ from $M$ to $\infty$ and by $t$ from $t_{z=3}$ (cosmological
time at $z=3$) to $t_0$.
We obtain
\begin{eqnarray}
& \ & \int_M^\infty M' \left( n(M',t_0) -n(M',t_{z=3}) \right) \ dM'
- \int_M^\infty \int_{t_{z=3}}^{t_0} M 
\frac{\partial n}{\partial t} \ dM' dt
\nonumber \\
& = & \int_M^\infty \int_{t_{z=3}}^{t_0} n(M',t)
      \langle \dot M' \rangle \ dM' dt    \nonumber \\
& \ &  +  \int_M^\infty \int_{t_{z=3}}^{t_0} (M'-M)
\left( \gamma_{\rm merge}(M',t)+\gamma_{\rm form}(M',t) \right)
\ dM' dt,  
\label{eq14} 
\end{eqnarray}
where $n(M,t_0)$ is the local mass function of SMBHs, $dN/dM(M,t_0)$,
as we defined in the previous section.
We substitute eq.(\ref{conti}) for the second term on the left-hand
side of eq.(\ref{eq14}), we obtain the equation,
\begin{equation}
F_{\rm local}(M)-F_{z=3}(M) - \widehat{F}_{\rm acc}(M)
= F_{\rm acc}(M) + F_{\rm merge + form}(M).
\end{equation}
Each terms is defined as below,
\begin{equation}
F_{\rm local}(M) \equiv
\int_M^\infty M' n(M',t_0) \ dM',
\end{equation}
\begin{equation}
F_{\rm z=3}(M) \equiv
\int_M^\infty M' n(M',t_{z=3}) \ dM',
\end{equation}
\begin{equation}
\widehat{F}_{\rm acc}(M) \equiv
\int_{t_{z=3}}^{t_0} M n(M,t) \langle \dot M \rangle \ dt ,
\end{equation}
\begin{equation}
F_{\rm acc}(M) \equiv
\int_M^\infty \int_{t_{z=3}}^{t_0} n(M',t)
      \langle \dot M' \rangle \ dM' dt ,
\end{equation}
\begin{equation}
F_{\rm merge + form}(M) \equiv
\int_M^\infty \int_{t_{z=3}}^{t_0} M' 
\left( \gamma_{\rm merge}(M',t) + \gamma_{\rm form}(M',t) \right)
\ dM' dt.
\end{equation}
We note that only SMBH-SMBH mergers, in which the mass of two SMBHs,
$M_1$ and $M_2$, satisfy the condition that 
$M_1 > M$ and $M_2 > M$, do not contribute to $F_{\rm merge + form}(M)$.
This is because these mergers do not change the mass density 
contained in SMBHs more massive than $M$. 
As discussed in the previous section, $F_{\rm merge + form}(M)$ is not negative, thus
we obtain the inequality,
\begin{equation}
F_{\rm z=3}(M) \leq F_{\rm local}(M) - F_{\rm acc}(M) 
- \widehat{F}_{\rm acc}(M).
\label{ineq2} 
\end{equation}

The physical meaning of each term is evident.
The cumulative mass density of SMBHs at $z=0$,
$F_{\rm local}(M)$, is calculated with the local mass function
given by AR02.
The accreted mass into SMBHs more massive than $M$ during
$0<z<3$, $F_{\rm acc}(M)$, is calculated by
\begin{equation} 
F_{\rm acc}(M)=
\int_{L(M)}^\infty dL' \int_{t(z=3)}^{t_0}
\frac{d\Phi(L',t)}{dL'} 
\frac{(1-\epsilon) L_{\rm bol}'}{\epsilon c^2} \ dt,
\end{equation}
which resembles $G_{\rm acc}(M,t_0)$ defined by eq.(\ref{gacc}).
Furthermore, the mass ``advection'' of SMBHs across the boundary mass $M$ 
(already mentioned in \S 2.1) by the mass accretion during $0<z<3$,
 $\widehat{F}_{\rm acc}(M)$, is calculated by
\begin{equation}
\widehat{F}_{\rm acc}(M)=
\frac{(1-\epsilon)L_{\rm bol}(M)}{\epsilon c^2}
\int_{t(z=3)}^{t_0} \frac{d\Phi(L,t)}{dL}
\frac{dL}{dM} M \ dt.
\end{equation}
With the inequality (\ref{ineq2}), we can estimate the
upper limit of the cumulative mass density of SMBHs at
$z=3$ ($F_{z=3}(M)$), $F_{\rm local}(M) - F_{\rm acc}(M) -
\widehat{F}_{\rm acc}(M)$.

\subsection{Tentative SMBHs in LBGs}

Here, we conjecture that tentative SMBHs exist in LBGs.
In the local universe, the BH mass is not correlated with 
the disk mass, but with the bulge mass of the galaxy. 
We note that we tentatively treat the SMBHs in LBGs, 
which has no clear bulge and disk components (or no clear 
distinction between these components). Therefore, we simply
assume that the mass of SMBHs is related to the stellar mass 
of LBGs as $M = \xi M_{\rm *}$, (see Steidel et al. 2002).
We treat the mass ratio, $\xi$, as a free parameter.
To compare the proportional relation for the normal galaxies
at $z=0$, $M/M_{\rm bulge} \sim 0.001-0.006$, we examine three cases, 
$\xi = 0.002,0.005$, and $0.01$. 
For the LBGs which have a typical stellar mass $(\sim 10^{10}M_\odot)$, 
the mass of corresponding tentative SMBHs is $10^7 - 10^8 M_\odot$.  
The cumulative mass density of
this tentative SMBHs in LBGs at $z=3$, $F_{\rm LBG}(M)$, is given by
\begin{equation}
F_{\rm LBG}(M) = 
\int_M^\infty 
\eta_{\rm LBG}(M')
\frac{d\phi(M_V)}{dM_V} \frac{dM_V}{dM_{\rm *}}
\frac{1}{\xi} \ dM',
\label{flbg}
\end{equation}
where $d\phi(M_V)/dM_V$ is the LFs of LBGs at $z=3$ in the rest-frame 
optical (V-band) luminosity and $\eta_{\rm LBG}(M)$ is the probability
that one LBG possesses a SMBH of mass $M$.
The LFs of LBGs are typical Schechter-type functions and we use the one
given by Shapley et al. (2001).
Shapley et al. also analyze about 100 LBGs with a population synthesis model, 
assuming a constant star formation rate. They
show the relation between the stellar mass of LBGs and
apparent luminosity in V-band (see their Fig.13), which is approximately
expressed as
\begin{equation}
M_V = -\frac65 \left( 
\log \left( \frac{M_{\rm *}}{M_\odot}h^2 \right) +8.0 \right)
+5.0 \log(h).
\end{equation}
Here, we define the mass average of $\eta_{\rm LBG}(M)$ as
\begin{equation}
\overline{\eta}_{\rm LBG}(M)
\equiv
\frac{F_{\rm LBG}(M)}
{\displaystyle{ 
\int_M^\infty \frac{d\phi(M_V)}{dM_V} \frac{dM_V}{dM_{\rm *}}
\frac{1}{\xi} \ dM } }
\equiv
\frac{F_{\rm LBG}(M)}{\widehat{F}_{\rm LBG}(M)}.
\end{equation}
We reliably set $F_{z=3}(M) \geq F_{\rm LBG}(M)$,
since there may be other populations of galaxies at $z \sim 3$ besides 
LBGs and these population may also possess the SMBHs.
Very recently, Franx et al.(2003) found the significant population
of red galaxies in $z>2$, which are chosen by their colors,
$J_s-K_s>2.3$. They point that the number density of these red
galaxies at $z \sim 3$ can be a half of that of LBGs. 
Furthermore, van Dokkum et al. (2003) argue that several among these
high-{\it z} red galaxies display the AGN-like rest-frame UV spectra.
More spectroscopy with more samples will make the AGN fraction
of these high-{\it z} red galaxies clear. 
With eq.(\ref{ineq2}), we can calculate the upper limit of
$\overline{\eta}_{\rm LBG}(M)$,  
\begin{equation}
\overline{\eta}_{\rm LBG}(M) \leq 
\frac{F_{\rm local}(M)-F_{\rm acc}(M)-\widehat{F}_{\rm acc}(M)}
     {\widehat{F}_{\rm LBG}(M)} 
\equiv \overline{\eta}_{\rm LBG, max}(M).
\label{eta}
\end{equation}
We again stress that this is a rather loose upper limit due to
some reasons. We simply subtract the accreted mass density during
$0<z<3$ from the mass density of SMBHs found in the center of local galaxies
and compare it with that of tentative SMBHs in LBGs.
If we can know the merger rates or direct formation rates
at each redshift, we can subtract $F_{\rm merge + form}(M)$
and obtain a more strict upper limit.
To investigate the redshift distributions of SMBH-SMBH merger rates
and direct formation rates of SMBHs, the gravitational wave (GW)) will be
a powerful tool. Today, several powerful GW detectors are under preparation,
for example, {\it Laser Interferometer Satellite Antenna} (LISA) in
the low frequency band ($10^{-4}-10^{-1}$ Hz) and {\it Laser
Interferometer Gravitational-Wave Observatory} (LIGO) in the high
frequency band ($10^1 - 10^3$ Hz).  
Since both the mergers between SMBHs and the gravitational collapse of SMS
will emit the low frequency GW (see Haehnelt 1994 for mergers, Saijo
et al. 2002 for collapse of SMS), these events are main targets of LISA.  
Furthermore, these any cosmological events at $z<10$ should be reliably
detected with high signal-to-noise ratio with LISA. Currently LISA is 
planned to start observations at the earliest in 2011.

\subsection{Upper Limit of the Probability that LBGs have SMBHs}

\subsubsection{Case (i): Non-negligible Contribution of Mergers and/or Direct Formation of SMBHs}

We calculate $\overline{\eta}_{\rm LBG,max}(M)$,
the upper limit of the mass averaged probability that
one LBG possesses a SMBH more massive than $M$, by eq.(\ref{eta}).
Of course, $\overline{\eta}_{\rm LBG,max}(M)$ depends on 
mass conversion efficiency, $\epsilon$, and Eddington ratio,
$f_{\rm Edd}$. First, we assume that these parameters are constants.
As discussed in the previous section, in this case, the contribution
of mergers and/or direct formation of SMBHs is not negligible and become
significant in the low-mass end.
Incidentally, many theoretical models for QSO LFs at high redshifts
based on the hierarchical structure formation adopted the constant
parameters (e.g. Hosokawa 2002, Wyithe \& Loeb 2002).
Of course, even if we can not neglect the mergers and/or direct
formation, the parameters do not have to be constant.
Here, we calculate $\overline{\eta}_{\rm LBG,max}(M)$ with constant
parameters as one possible case with non-negligible mergers and/or
direct formation of SMBHs. 
As we have shown in the previous section, we cannot choose these
parameters freely since the inequality (\ref{ineq}) must
be satisfied. 
We calculate $\overline{\eta}_{\rm LBG,max}(M)$ with the parameter sets, 
$(\epsilon, f_{\rm LBG})$, indicated by points 1 $\sim$ 3 in Fig.2.
Below, we choose the mass range
of tentative SMBHs to be $10^7 M_\odot < M < 10^{8.5} M_\odot$
to calculate $\overline{\eta}_{\rm LBG,max}(M)$. 
For $M<10^7 M_\odot$, the luminosity range of LBGs
corresponds to the faint end of LFs at $z \sim 3$, thus no sufficient data is available.
The fitting formulae for LFs of Shapley et al. (2001)
are valid only for $M_V-5 \log(h) < -20.5$.
Whereas, the HXLFs of AGNs given by UAOM03 are valid to the 
sufficiently faint end, $\log L_x = 41.5$, which corresponds to 
$M \sim 10^5 M_\odot$ if we assume $f_{\rm Edd}=1.0$. 
For $M>10^{8.5}M_\odot$, conversely, it is difficult to satisfy
the inequality eq.(\ref{ineq}).

\begin{figure}
\plotone{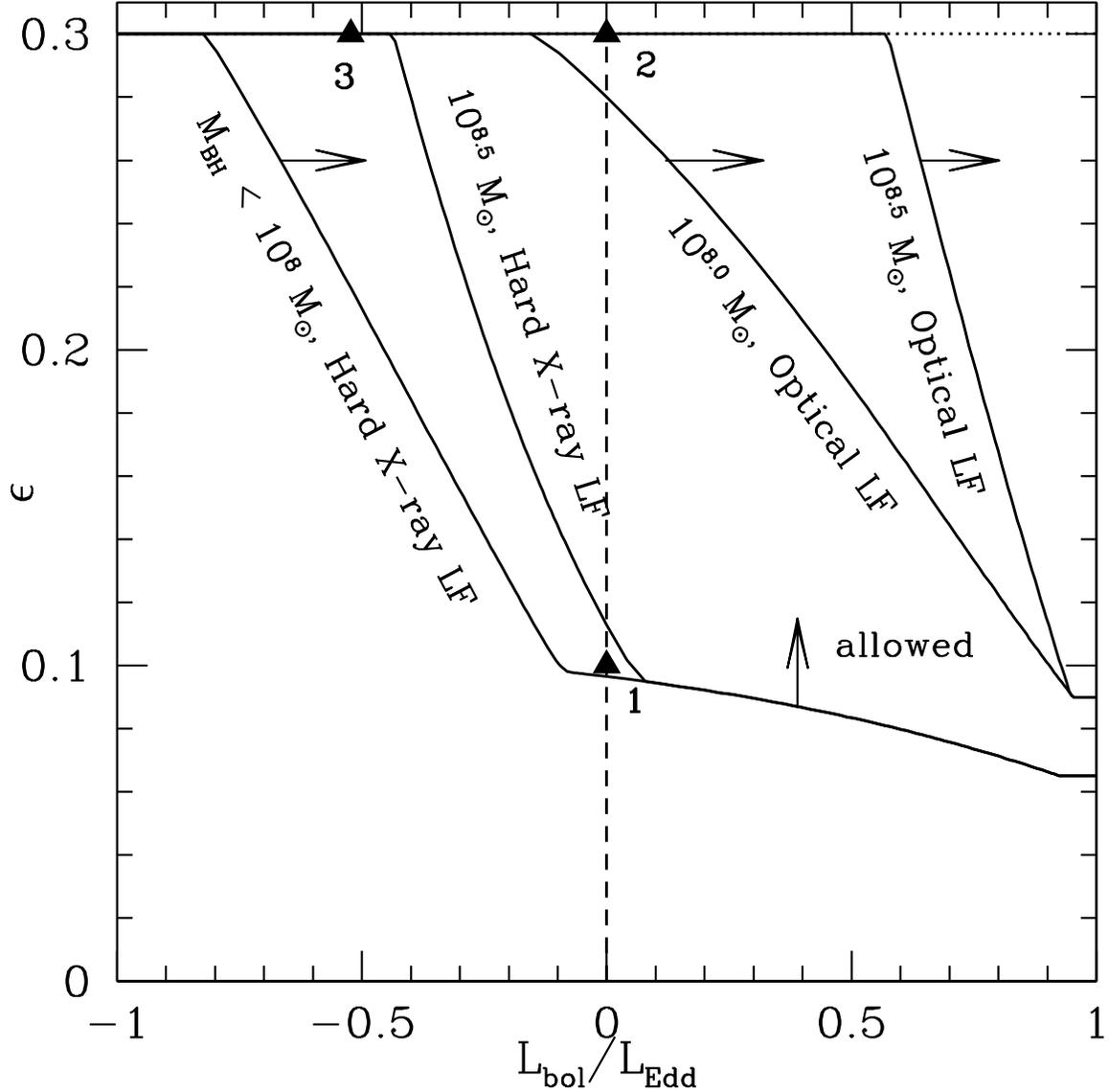}
\figurenum{2}
\label{fig.search}
\figcaption{The region of the parameter space where the inequality,
$G_{\rm local}(M,t_0) \geq G_{\rm acc}(M,t_0)$ is satisfied only at
$M \leq M_0$. 
We calculate $G_{\rm local}(M,t_0)$ using the local mass function 
of SMBHs (a) (see text in detail).  
The allowed region for each $M_0$ is above the labeled lines
for $M_0 = 10^{8.0}$ and  
$10^{8.5} M_\odot$. We set the upper limit of the energy conversion
efficiency as $\epsilon=0.3$. The representative parameter sets
are indicated by points 1 $\sim$ 3. We can show the corresponding
$G_{\rm acc}(M,t_0)$ for 1 $\sim$ 3 in Fig.\ref{fig.gcalc}.}
\end{figure}

\begin{figure}
\plotone{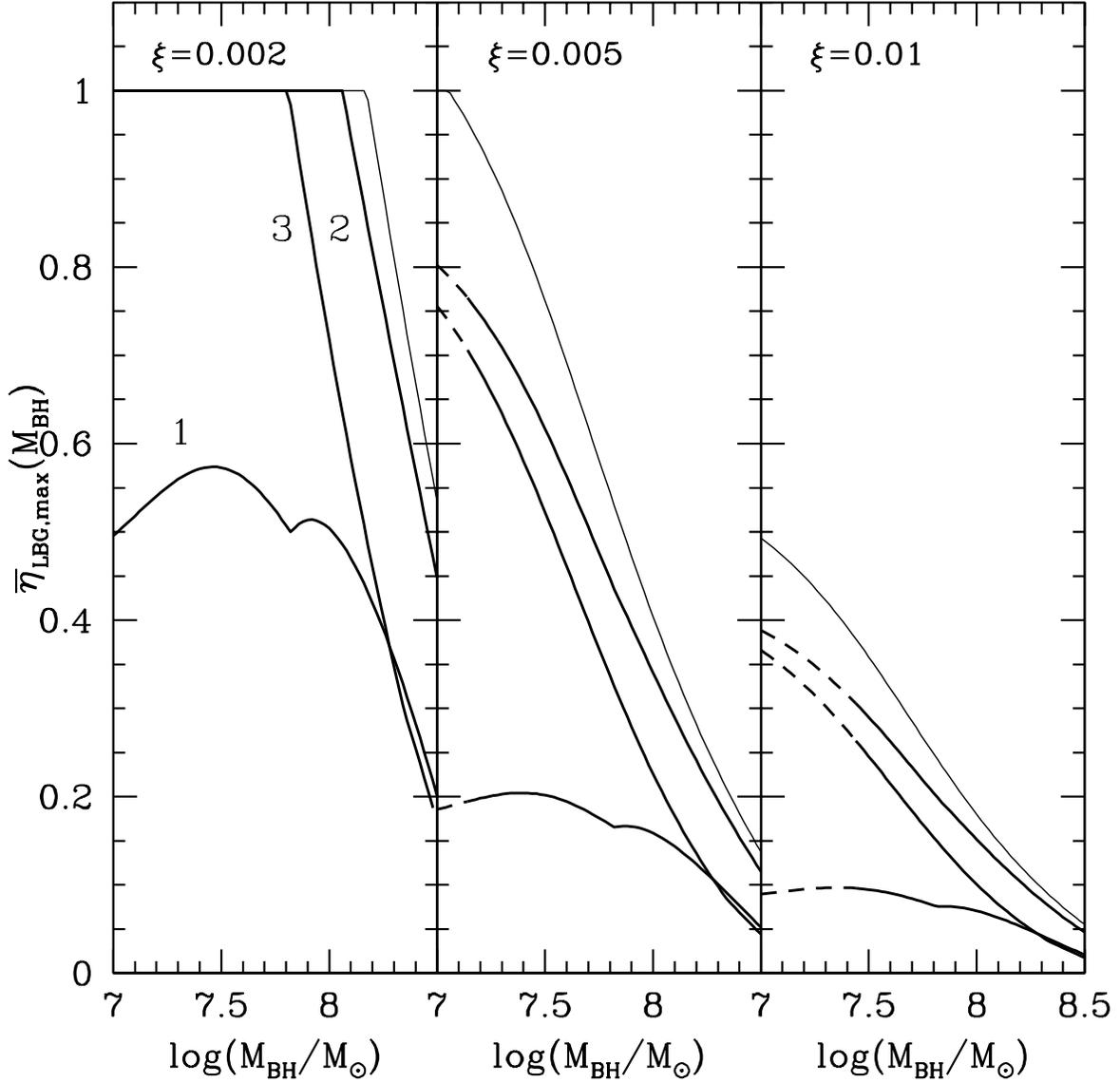}
\figurenum{3}
\label{fig.comp}
\figcaption{Mass averaged probability that one LBG
possess a SMBHs more massive than $M$, $\overline{\eta}_{\rm LBG,max}(M)$
for $\xi=0.002, 0.005$ and 0.01.
The lines represent 
$\overline{\eta}_{\rm LBG,max}(M)$ calculated with parameter sets
indicated in Fig.\ref{fig.search} by points 1 $\sim$ 3 
(left to right; point 1, 3, 2 in turn).
In the figure, dashed lines represents the mass range where
the luminosity of the corresponding LBG is too faint : $M_{\rm V}-5 \log(h)
> -20.5$.
The thin solid lines represents 
$F_{\rm local}(M)/\widehat{F}_{\rm LBG}(M)$.}
\end{figure}

We represent the calculated $\overline{\eta}_{\rm LBG,max}(M)$ 
for $\xi = 0.002$, 0.005 and 0.01 in Fig.\ref{fig.comp} 
with SMBH mass function (a). 
In these figures dashed lines represent the mass range where
the luminosity of the corresponding LBG is too faint, $M_{\rm V}-5 \log(h)
> -20.5$. We can see that $\overline{\eta}_{\rm LBG,max}(M) \sim
1$ is possible for $\xi = 0.002$ with large energy conversion efficiency
($\epsilon \sim 0.3$) (points 2 and 3 in Fig.\ref{fig.search})
; that is, almost all LBGs can possess the
SMBHs. This value, $\xi = 0.002$, is the same as the ratio between the 
bulge mass and the SMBH mass found with local galaxies.
However, for $\xi = 0.01$ only a part of LBGs can possess the SMBHs ;
the upper limits are about 30\%. 
If we use mass function (b), calculated $\overline{\eta}_{\rm
LBG,max}(M)$ becomes larger only by factor 1.5 than that in
Fig.\ref{fig.comp} with similar parameter sets.
We can explain why $\overline{\eta}_{\rm LBG,max}(M)$
increases steeply as $M$ decreases. As mentioned in the previous section,
the contribution of mergers and/or direct formation 
to grow SMBHs increases as $M$ decreases with the constant
parameters of mass accretion. We do not include mergers and/or direct
formation of SMBHs ($F_{\rm merge + form}(M)$) in eq.(\ref{eta})
since the event rates of these processes are highly unknown,
then the constraint on $\overline{\eta}_{\rm LBG}(M)$ becomes loose
as $M$ decrease. 
Therefore, if we can know the redshift distribution of merger rates and/or
direct formation rates of SMBHs, we can make the upper limit more
strict. Especially, if the mergers and/or direct formation occur mainly
at $z<3$, the constraint will be as strict as that with point 1 
in Fig.\ref{fig.search}.
In Fig.\ref{fig.comp}, we also plot $F_{\rm local}(M)/\widehat{F}_{\rm LBG}(M)$
for comparison. This is the upper limit for which the population
of tentative SMBHs do not exceed the local population of SMBHs.

\begin{figure}
\figurenum{4}
\label{fig.tm_all}
\plotone{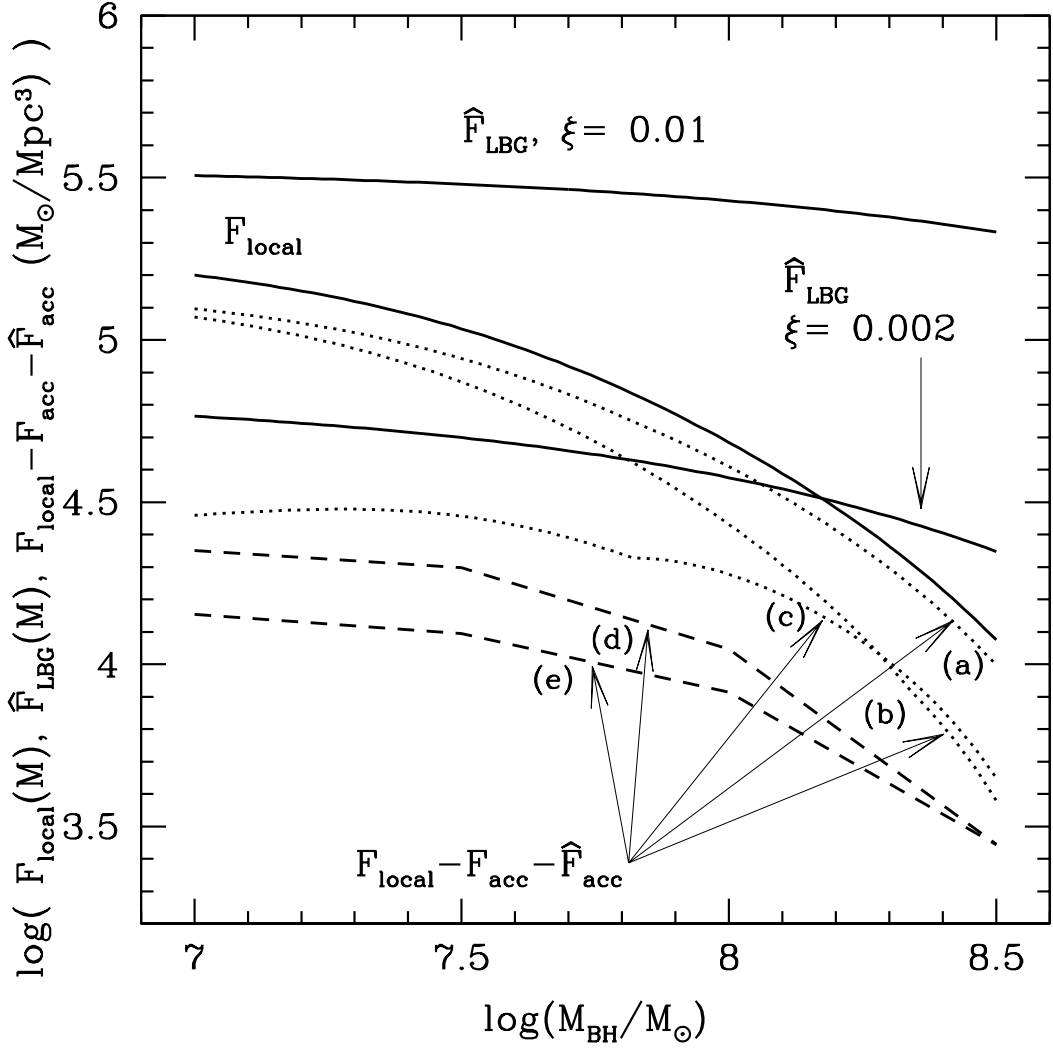}
\figcaption{
The mass dependence of the cumulative mass densities of SMBHs;
$F_{\rm local}(M)$: observational estimate at $z=0$, 
$\widehat{F}_{\rm LBG}(M)$: assuming that all LBGs have SMBHs with 
mass ratio, $\xi=0.01$ and 0.002, at $z=3$,
$F_{\rm local}(M)-F_{\rm acc}(M)-\widehat{F}_{\rm acc}(M)$: the upper
 limit at $z=3$.
We represent $F_{\rm local}(M)-F_{\rm acc}(M)-\widehat{F}_{\rm acc}(M)$
in five cases; (a): with constant parameters indicated by point 2 in Fig.2
, (b): same as (a) but for point 3, (c): same as (a) but for point 1,
(d): with mass dependent energy conversion efficiency (eq.(\ref{eps_m})) and
$f_{\rm Edd}=1.0$, (e): same as (d) but for $f_{\rm Edd}=0.3$.}
\end{figure}

Fig.\ref{fig.tm_all} represents the mass dependence of 
$F_{\rm local}(M)$, $\widehat{F}_{\rm LBG}(M)$ and
$F_{\rm local}(M)-F_{\rm acc}(M)-\widehat{F}_{\rm acc}(M)$ respectively.
These are all cumulative mass densities; $F_{\rm local}(M)$ is the
observational estimate at $z=0$, $\widehat{F}_{\rm LBG}(M)$ is one
assuming that all LBGs have SMBHs with mass ratio, $\xi$, and
$F_{\rm local}(M)-F_{\rm acc}(M)-\widehat{F}_{\rm acc}(M)$ is the upper
limit at $z=3$.
In this figure, we use the mass function (a) and
adopt the parameter set displayed with the points 1-3 in Fig.\ref{fig.search}.
It is clear that if all LBGs have SMBHs with large mass ratio as $\xi =
0.01$, the resultant mass function exceed even the local estimate
($F_{\rm local}(M)$). 
The ratio between $\widehat{F}_{\rm LBG}(M)$ and 
$F_{\rm local}(M)-F_{\rm acc}(M)-\widehat{F}_{\rm acc}(M)$
is our calculating $\overline{\eta}_{\rm LBG,max}(M)$.
 
We must note that we can only set the {\it upper} limit
of the probability, in principle, that LBGs have SMBHs.
Then, we can not even rule out that any normal LBGs (not AGNs) do not have
SMBHs. Here, we conclude as follows.
If we assume the mass ratio, $\xi ( \equiv M_{\rm BH}/M_*)$, of tentative SMBHs
in LBGs is comparable to be observed $M_{\rm BH}/M_{\rm bulge}$ of
normal galaxies at $z=0$, 
we cannot completely rule out that the large part of LBGs 
possess inactive SMBHs. 
However, some conditions are needed in this case.
That is, large energy conversion efficiency ($\epsilon \sim 0.3$)
and significant mergers and/or direct formation of SMBHs at high
redshifts ($z>3$).
Whereas, if $\xi$ is larger than the local
value of $M_{\rm BH}/M_{\rm bulge}$, the fraction of LBGs which can possess
SMBHs decrease to less than several tens of \%.

\subsubsection{Case (ii): Dominant Contribution by Mass 
Accretion to Grow SMBHs} 

Some recent works suggest that the mass accretion may be the
dominant process to grow SMBHs 
( Yu \& Lu 2004, Marconi et al. 2004).
Here, we consider the special case, where the mergers and direct
formation of SMBHs through gravitational collapse of the SMS are not important
for $M \geq 10^7 M_\odot$.
Concretely, $G_{\rm merge + form}(M,t_0)$ can be neglected and
the equality is satisfied in eq.(\ref{ineq}),
\begin{equation}
G_{\rm local}(M,t_0)=G_{\rm acc}(M,t_0).
\label{equali}
\end{equation}
In this case, we differentiate eq.(\ref{equali}) by M and we can represent
the energy conversion efficiency as the function of $M$, 
\begin{equation}
\frac{\epsilon}{1-\epsilon}(M)
= \frac{
\displaystyle{\frac{L_{\rm bol}(M)}{c^2} 
              \int_0^{t_0} \frac{d\Phi}{dL}( L(M),t) \frac{dL}{dM} \ dt }}
{\displaystyle{
 \int_M^\infty n(M',t_0) \ dM'}},
\label{eps_m}
\end{equation}
(YT02). Inversely, if one calculate $G_{\rm local}(M,t_0)$ and $G_{\rm
acc}(M,t_0)$ using $\epsilon(M)$ with eq.(\ref{eps_m}),
eq.(\ref{equali}) is naturally satisfied in all mass range.
The functional form of $\epsilon(M)$ is the same as that in Fig.4. of YT02. 
Here, we calculate $\overline{\eta}_{\rm LBG,max}(M)$ with the mass
dependent energy conversion efficiency, $\epsilon(M)$ given by
eq.(\ref{eps_m}). We note that when mergers and direct formation of
SMBHs are negligible, the equality is satisfied in eq.(\ref{ineq2}).
However, the inequality $F_{z=3}(M) \geq F_{\rm LBG}(M)$ still
remains due to possible other populations of hosts of SMBHs besides LBGs.
Therefore, $\overline{\eta}_{\rm LBG,max}(M)$ calculated with
$\epsilon(M)$ still gives the upper limit of 
$\overline{\eta}_{\rm LBG}(M)$ though it is more stringent than
the case with constant parameters. 
 
\begin{figure}
\figurenum{5}
\label{fig.comp_eps}
\plotone{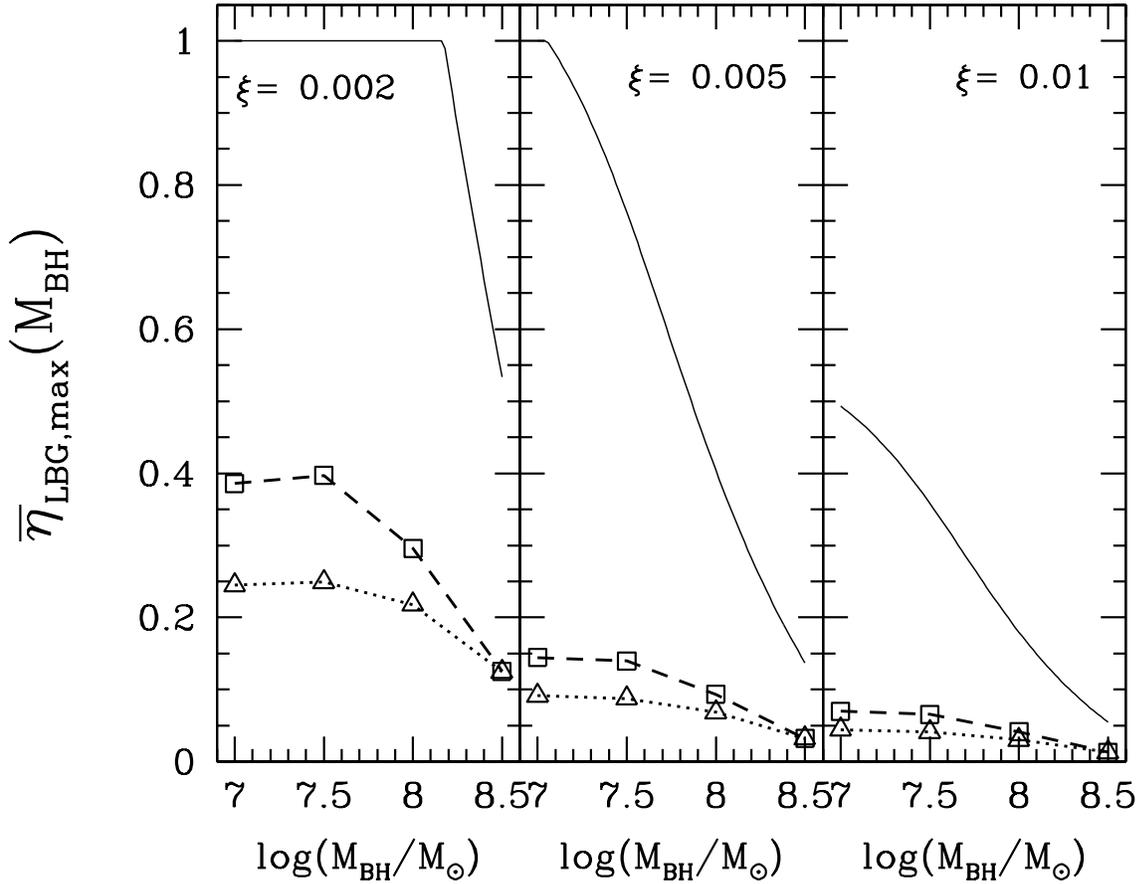}
\figcaption{
Mass averaged probability that one LBG
possess a SMBHs more massive than $M$, $\overline{\eta}_{\rm LBG,max}(M)$
with the mass dependent energy conversion efficiency given by
 eq.(\ref{eps_m}) for $\xi = 0.002$, 0.005 and 0.01 respectively. Dashed lines 
and dotted lines represents the results with $f_{\rm Edd}=1.0$ and $0.3$
respectively.  
We calculate $\overline{\eta}_{\rm LBG,max}(M)$ only at 
$M=10^7, 10^{7.5}, 10^{8}$ and $10^{8.5} M_\odot$ and connect these
values with lines. The thin black solid lines represents 
$F_{\rm local}(M)/\widehat{F}_{\rm LBG}(M)$.}
\end{figure}

Fig.\ref{fig.comp_eps} represents the results. As expected, the calculated
$\overline{\eta}_{\rm LBG,max}(M)$ is more strict than the results with
constant parameters (Fig.\ref{fig.comp}). Even if the value of $\xi$ is
comparable to $M_{\rm BH}/M_{\rm bulge}$ (0.001-0.006) of local
galaxies, only a part of LBGs are allowed to have SMBHs 
(less than 10-30\%). 
Especially, if $\xi$ is larger than $M_{\rm BH}/M_{\rm bulge}$,
LBGs which possess SMBHs become very rare (less than 10 \%).
If this is the case, 3\% of LBGs which show the AGN activity may be the
special LBGs where SMBHs are born within the galaxies. 
In Fig.\ref{fig.comp_eps}, we also plot 
$F_{\rm local}(M)/\widehat{F}_{\rm LBG}(M)$ for reference. 
These lines are same as Fig.\ref{fig.comp}. In this case, 
the difference between calculated $\overline{\eta}_{\rm LBG,max}(M)$ and 
$F_{\rm local}(M)/\widehat{F}_{\rm LBG}(M)$ is large. 
In Fig.\ref{fig.tm_all}, we can see that 
$F_{\rm local}(M)-F_{\rm acc}(M) -\widehat{F}_{\rm acc}(M)$ is
smaller than $F_{\rm local}(M)$ by about one order of magnitude
in $10^{7} M_\odot < M < 10^{8.5} M_\odot$.

\subsubsection{Necessary Condition to Neglect Mergers and/or Direct Formation}

As we show above, how much mergers and direct formation of SMBHs occur
until $z=0$ is important to estimate the mass function of
SMBHs at high redshifts ($z \sim 3$ in this paper). 
As mentioned above, we can observe these events by the gravitational
radiation with LISA in the future.
Here, we crudely estimate the necessary condition to drop the merger
term in eq.(\ref{geq}).
In eq.(\ref{mg}) $\Omega(M_1,M_2)$ is the number of merger events among the SMBHs
of mass $M_1$ and $M_2$ per unit volume per unit cosmological time. 
We can express this $\Omega(M_1,M_2)$ as,
\begin{equation}
\Omega(M_1,M_2)= 
\frac{\Delta N}{\Delta M_1 \Delta M_2 \Delta t \Delta V}.
\end{equation}
With this expression, the contribution of mergers to 
$G_{\rm merge + form}(M,t_0)$ may be written as
\begin{equation}
G_{\rm merge}(M,t_0) \equiv
\int_M^\infty \int_0^{t_0} f(M,M')\gamma_{\rm merge}(M',t) \ dM'dt
\sim
\frac{M' \Delta N}{\Delta \ln M \Delta \ln M' \Delta z \Delta V}
\cdot \Delta z
\end{equation}
Here, we transform the time interval, $\Delta t$ to the redshift
interval $\Delta z$. If $G_{\rm merge}(M,t_0)$ is less than
$G_{\rm local}(M,t_0)$ and $G_{\rm acc}(M,t_0)$, say, 
at $M \sim 10^7 M_\odot$, we have
\begin{equation}
\frac{\Delta N}{\Delta \ln M \Delta \ln M' \Delta z \Delta V}
\cdot \Delta z
\leq 10^{-2} (\mbox{Mpc}^{-3}) ,
\label{nmerg}
\end{equation}
where we use $G_{\rm local} \sim G_{\rm acc} \sim
10^5 M_\odot \cdot \mbox{Mpc}^{-3}$ at $M \sim 10^7 M_\odot$ (see Fig.1)
and we set $M' \sim 10^7 M_\odot$.
Of course, other mergers than $M \sim M' \sim 10^7 M_\odot$ contribute
to $G_{\rm merge}(10^7 M_\odot, t_0)$, but we estimate loose upper limit
of the event rate of mergers where $M \sim M' \sim 10^7 M_\odot$ to drop the
merger term in eq.(\ref{geq}). 
We can convert eq.(\ref{nmerg}) to
the number of the events which we can observe per unit time
from the earth,  
\begin{equation}
\frac{dN}{d \ln M d \ln M' dz dt} \cdot dz
= \frac{\Delta N}{\Delta \ln M \Delta \ln M' \Delta z \Delta V}
  \cdot \Delta z
  \times 4 \pi c (1+z) \cdot d_A^2(z),
\end{equation}
where $d_A(z)$ is the angular diameter distance. The time interval $dt$
on the left-hand side is related to the volume interval $\Delta V$ on
the right-hand side as $\Delta V = 4 \pi c (1+z) d_A^2 dt$.
If we set $z=3$, the angular diameter distance is
$d_A(z=3) \sim 1 \mbox{Gpc}$ in the $\Lambda$CDM universe (the redshift
dependence of $d_A$ is not large), then we have
\begin{equation}
\frac{dN}{d \ln M d \ln M' dz dt} \cdot dz
\leq 0.6  \ \ (\mbox{times} \cdot {\rm yr}^{-1}) 
\end{equation}
with eq.(\ref{nmerg}). This is a necessary condition for event rate
of mergers of $M \sim M' \sim 10^7 M_\odot$ to drop 
$G_{\rm merge}(10^7 M_\odot,t_0)$ in eq.(\ref{geq}). 
That is, if we observe one or more events
of the SMBH-SMBH merger between $10^7 M_\odot$ SMBHs per one year, 
the effect of mergers to grow SMBHs is at least not negligible,
compared with the mass accretion. 
To neglect the mergers in terms of the continuity equation of SMBHs,
no detection of events for several years is at least needed.
We can determine the contribution of mergers observationally in the
future.

\section{Discussions}

\subsection{Scatter in the $M-\sigma$ Relation}

In \S2.2, we have searched the parameter sets, $(\epsilon, f_{\rm
Edd})$  which satisfy inequality (9) with the HXLFs of AGNs (Fig.2).
In fig.2, we can find the allowed region of the parameter
space where ineq.(9) is satisfied only in $M \leq 10^{8.5} M_\odot$.
However, if we demand sub-Eddington case, $f_{\rm Edd} \leq 1.0$, 
it is still difficult to satisfy ineq.(9) in the high-mass end 
($M \geq 10^9 M_\odot$). For example, even if we adopt the most
adequate parameter set, $(\epsilon, f_{\rm Edd})=(0.3,1.0)$ to
suppress $G_{\rm acc}(M,t_0)$, ineq.(9) is satisfied only in
$M \leq 10^9 M_\odot$ (see Fig.1). 
One possible resolution of this discrepancy is the scatter 
included in the mass function of the SMBHs. That is,
the scatter in the $M-\sigma$ relation. Fortunately, Yu \& Lu (2004)
study the effect of the scatter on the mass function. 
They assume the Gaussian scatter for the
distribution of $\log M_{\rm BH}$ at a given $\sigma$ (see their
eq.(43)), and find that only the high-mass end of the mass function
$(>3 \times 10^8 M_\odot)$ is sensitive to the scatter. 
Furthermore, the scatter (they test the 3 cases for the intrinsic 
standard deviation, $\Delta_{\log M_{\rm BH}}=0,0.27$ and 0.4) 
{\it increase} the high-mass end (see their
Fig.1). Especially, they show that the effect of the scatter is 
significant in $M \geq 10^9 M_\odot$. This is an important result.
The discrepancy discussed above will be resolved with the accurate
decision of the intrinsic scatter in the $M-\sigma$ relation.
Incidentally, we note that the fitting function of the mass function 
by AR02 used in this paper becomes worse in the high-mass end 
($M > 10^{8.5} M_\odot$) (see their Fig.8 and 9).

\subsection{Relaxing Assumptions}

In this paper, we use some assumptions for simplicity. Here, we discuss
the effects of relaxing these assumptions.
Some of these assumptions are listed as following,

\begin{itemize}
\item[(i)] extrapolation of HXLFs of AGNs to $z>3$,

\item[(ii)] no redshift evolution of $\epsilon$ and $f_{\rm Edd}$,

\item[(iii)] mass conservation during mergers.

\end{itemize}

Throughout this paper,
we extrapolate the observed HXLFs which are valid
only at $z<3$ to the higher redshifts (assumption (i)). 
However, we note that the calculations in this paper 
depend only on the HXLFs just before $z=3$. 
To confirm this, we perform the same 
calculations with the tentative HXLFs,
$d\Phi(L_{\rm X},z)/d\log L_{\rm X} = 0$ for $z>3.5$. 
First, the critical value of $\epsilon$ for a given $f_{\rm Edd}$
to satisfy ineq.(\ref{ineq}) becomes less than 15\% smaller than
that of Fig.2.  
Next, $\overline{\eta}_{\rm LBG, max}(M)$ becomes less than 5\% smaller
than that of Fig.3. Therefore, we conclude that the results
in this paper are robust unless the HXLFs change suddenly
from the extrapolation of the observed ones at $z<3$ just before $z=3$. 

Since the redshift evolution of the parameters, $\epsilon$ and
$f_{\rm Edd}$ is uncertain, we adopt the assumption (ii).
Due to this assumption, we avoid to introduce extra parameters
to represent the redshift evolution.  For the energy conversion
efficiency, $\epsilon$, we have no idea on the redshift evolution.
Many theoretical works adopt $\epsilon$ with no redshift
evolution (e.g. Kauffmann \& Haehnelt 2000, Wyithe \& Loeb 2002, 
Volonteri, Haardt \& Madau 2003). For Eddington ratio, $f_{\rm
Edd}$, several theoretical works argue that $f_{\rm Edd}$ decreases 
as redshift decreases and that this is one possible origin of the rapid decline
in population of bright QSOs from $z \sim 2.5$ to $z = 0$
(e.g. Kauffmann \& Haehnelt 2000, Haiman \& Menou 2000).
Therefore, we test the following simple decline of $f_{\rm Edd}$ to
relax assumption (ii),
\begin{equation}
f_{\rm Edd}(z) = \left\{
              \begin{array}{lr}
              1.0,& \quad \mbox{for $z>2.5$} \\
              \frac{1.0-f_{\rm Edd,0}}{2.5} z + f_{\rm Edd,0},
              & \quad \mbox{for $z\leq2.5$}
              \end{array}\right.
\end{equation}
We calculate $G_{\rm acc}(M,t_0)$ for $f_{\rm Edd,0}=0.5,0.1,0.05$ and
0.01. However, the calculated $G_{\rm acc}(M,t_0)$ with $f_{\rm Edd}(z)$
is different from that with constant $f_{\rm Edd}$ only by less than
10\%. In this case, the best-fit values of constant $f_{\rm Edd}$
are $f_{\rm Edd}=0.8,0.64,0.62 \ (\mbox{corresponding to} \ 
f_{\rm Edd}(z=1.5))$ and $0.5 \ (f_{\rm Edd}(z=1.25))$ for 
$f_{\rm Edd,0}=0.5,0.1,0.05$ and 0.01 respectively.

During the merger of the SMBHs, the strong GW is emitted. 
Actually, the total mass of the SMBH binary
can generally decrease due to the GW emission during the merger
despite the assumption (iii).
For the most efficient radiative merging, the total entropy
(or the area) of SMBHs is conserved. In this case, the resultant mass
of the merger between SMBHs of $M_1$ and $M_2$ is $M = \sqrt{M_1^2 +
M_2^2}$. That is, about 30\% of $M_1 + M_2$ can be maximally 
radiated as the GW. 
In this paper, we constrain the upper bound of the population of SMBHs at 
$z=3$ based on the assumption that the mass density of SMBHs never
decreases. If significant mergers decrease the mass of SMBHs,
we may not rule out that there are already much SMBHs at $z=3$ than
that at $z=0$. 
YT02 consider the growth history of SMBHs for the maximally efficient 
radiative merging using the continuity equation. 
Here, we only note that the formulation in this paper can be applied 
for this case by replacing the variable, $M$ with the entropy (or area)
of the SMBHs, though there is the uncertainty on the spin parameter, $a$.

\subsection{Other Possible Constraint}

The current concrete observational information on the ``active'' LBGs
is only the number fraction to the all LBGs, $3\%$ ( Steidel et al. 2002,
Nandra et al. 2002). 
However, we can possibly use this fraction, $3\%$ as the
lower limit of $\overline{\eta}_{\rm LBG, max}(M)$. 
Though $3\%$ is much smaller than the calculated values of 
$\overline{\eta}_{\rm LBG, max}(M)$ in many cases, if this fraction
increases in the future, we can get the more strict lower limit.
For the dominant mass accretion case (\S3.3.2 or Fig.5), especially,
calculated $\overline{\eta}_{\rm LBG, max}(M)$ is comparatively small,
then the lower limit will be valid.
If we can put the meaningful lower limit as noted above, 
we can constrain the lower bound of Eddington ratio, 
$f_{\rm Edd}$ and the upper
bound of the mass ratio, $\xi$. In all cases, calculated 
$\overline{\eta}_{\rm LBG, max}(M)$ decreases to the high-mass end.
However, we must consider the scatter of the $M-\sigma$ relation
in the high-mass end to apply the lower limit (\S4.1).

\section{Conclusions}

In this paper, we study demographics of SMBHs in the local
and high-{\it z} universe and we crudely constrain the possibility that
LBGs have SMBHs for an application. 
To constrain the possible population of SMBHs within the LBGs,
we estimate the upper limit of mass density of SMBHs at $z \sim 3$.      
In other words, if there are too many and too massive SMBHs already at
$z \sim 3$, the resultant mass density of SMBHs at $z \sim 0$ 
exceed the local mass density of SMBHs.
We use the continuity equation of SMBHs, eq.(\ref{conti}) and
transform it to the useful integrated form.
We consider the three physical processes to grow the SMBHs;
mass accretion, mergers and direct formation of SMBHs.
Of these processes to grow SMBHs, we can use the HXLFs of AGNs to estimate the 
cosmological accretion history into the SMBHs.

First, we investigate the demographics of SMBHs in the local universe.
YT02 argue that inequality (\ref{ineq}) is not satisfied with the
accreted mass at $z>0$ estimated with OLFs of QSOs and with the local
mass function of SMBHs with typical parameter set, 
$(\epsilon,f_{\rm Edd})=(0.1,1.0)$. 
We use the HXLFs of AGNs including optically
absorbed AGNs, then the accreted mass with HXLFs should have become
larger than that with OLFs. However, we show that the discrepancy
is resolved using HXLFs with the same parameters and constant
$L_{\rm X}/L_{\rm bol}$.  
The origin of the discrepancy is probably the assumption of the constant
$L_{\rm B}/L_{\rm bol}$. The comparison between the redshift evolution
of OLFs and that of HXLFs suggests that $L_{\rm B} \propto L_{\rm
X}^{1.4}$. Therefore, if the assumption of the constant 
$L_{\rm X}/L_{\rm bol}$ is reasonable, the ratio of B-band luminosity
to the bolometric luminosity has the dependence as 
$L_{\rm B}/L_{\rm bol} \propto L_{\rm bol}^{-0.4} \propto M^{-0.4}$. 
This makes the redshift evolution of OLFs as slow as that of HXLFs
as $z$ increases, and calculated $G_{\rm acc}(M,t_0)$ with OLFs
becomes small.
Furthermore, we constrain the energy conversion efficiency, $\epsilon$ 
and Eddington rate, $f_{\rm Edd}$ to satisfy the inequality 
(\ref{ineq}) upto the high mass end 
($M < 10^{8} M_\odot$, $10^{8.5} M_\odot$). 
If we demand $f_{\rm Edd} \leq 1$, the allowed region of 
$(\epsilon,f_{\rm Edd})$ exist only if we use HXLFs 
($\epsilon \sim 0.1-0.3$ and $f_{\rm Edd} \sim 0.3-1.0$). 
Totally, we conclude on the demographics of SMBHs
in the local universe as follows.
\begin{itemize}
\item   The cosmological accretion history based on HXLFs is more
	plausible than that based on OLFs. This is because we can satisfy
        ineq.(\ref{ineq}) with typical parameters 
        (especially $f_{\rm Edd} \leq 1$). The reason why calculated 
        $G_{\rm acc}(M,t_0)$ with HXLFs is smaller than that with OLFs
        is that we assume constant luminosity ratio, $C_{\rm B}$ and
	$C_{\rm X}$ for both bands. 
\end{itemize}
With these constant parameter sets,
$G_{\rm local}(M,t_0) - G_{\rm acc}(M,t_0)$ becomes large at the low mass
end. This means that the contribution of mergers and/or direct formation
of SMBHs is important in the low mass end, if the physical
parameters are actually constant as we assumed.
We adopt the constant parameters to estimate the accreted mass to SMBHs
during $0<z<3$ for the case that there is non-negligible mergers and/or
direct formation in the latter-half of this paper.  

Next, we derive the inequality (\ref{ineq2}) between the population of SMBHs
at $z=0$ and $z=3$ and the accreted mass into SMBHs during $0<z<3$
from eq.(\ref{conti}) again. This gives the upper limit of
the cumulative mass density of SMBHs at any high redshifts.
For an application, we consider the tentative SMBHs within the LBGs.
We assume that the mass of SMBHs, $M$ is proportional to the stellar
mass of LBGs, $M_*$ with the parameter $\xi \equiv M/M_*$.
We define the mass averaged probability that one LBG harbors
a SMBH more massive than $M$, $\overline{\eta}_{\rm LBG}(M)$,
and calculate the upper limit of $\overline{\eta}_{\rm LBG}(M)$,
 $\overline{\eta}_{\rm LBG,max}(M)$.
We simply estimate the upper limit of cumulative mass density of SMBHs
at $z \sim 3$ considering only mass accretion during $0<z<3$.
For the event rate of mergers and/or direct formation of SMBHs, 
we consider two simple limit; (i) there are
non-negligible contribution of mergers and/or direct formation
(ii) the mass accretion is the dominant process to grow the SMBHs.
The constraint on the population of the tentative SMBHs in LBGs are 
summarized as follows.

\begin{itemize}
\item
In case (i), we adopt the constant parameters for one possible case
as mentioned above. 
If the assumed mass ratio, $\xi$ is comparable to 
$M/M_{\rm bulge}$ observed in the local galaxies $(\xi = 0.002)$, 
we cannot rule out that almost all LBGs can have SMBHs 
$(\overline{\eta}_{\rm LBG,max}(M) \sim  1.0 )$.
However, if this is the case, the large energy conversion efficiency,
$\epsilon \sim 0.3$ and significant mergers and/or direct formation
of SMBHs at $z>3$ will be needed.  
If $\xi$ is larger than the local value $(\xi > 0.005)$, only a part
of LBGs can have SMBHs ($\overline{\eta}_{\rm LBG,max}(M) \sim$  several
ten \%). 
\item
In case (ii), we can make the upper limit more strict.
In this case, we can express the energy conversion efficiency as
the function of the SMBH mass, $\epsilon(M)$. With this $\epsilon(M)$,
calculated $\overline{\eta}_{\rm LBG,max}(M)$ is only ten \% even if
$\xi$ is small. 
Especially, if $\xi$ is larger than the $M/M_{\rm
bulge}$ of local galaxies, $\overline{\eta}_{\rm LBG,max}(M)$ can become
less than 10\%. 
\end{itemize}
We roughly estimate the necessary condition to neglect the merger term
in eq.(\ref{geq}). For example, for the major merger between SMBHs of $10^7
M_\odot$, the loose upper limit is about 1 event per one year.
If we observe these merger events one or more times per year,
we can not neglect the SMBH-SMBH mergers. 
Totally, the possibility that nearly all LBGs have SMBHs with large
mass ratio, $\xi > 0.005$, is reliably ruled out. 
Even if AGN-like LBGs have SMBHs with large mass ratio as Vestergaard
(2003) and Vestergaard (2004)
suggested, remaining normal LBGs will not have inactive SMBHs with the same
mass ratio. 

Though we consider the tentative SMBHs in LBGs with mass ratio,
$\xi \equiv M/M_*$, this proportional relation is only a
trial one. Papovich et al.(2001) note that the stellar mass
estimated with the rest-frame UV-to-optical spectrum is only
the mass of young stars. That is, additional population of old stars
may exist. Papovich et al. estimate that the upper limit of the mass
of old stars is 3-8 times as large as that of young stars. 
Since the SMBH mass is not correlated with the younger disk component
but older bulge component in the nearby galaxies, it may be better
to relate the tentative SMBH mass with the mass of old stars
if they exist. However, our approach will be still valid,
since we simply try another values of $\xi$. 
Of course, even if SMBHs actually exist in the high-z galaxies, the 
proportional relation between $M$ and $M_{\rm bulge}$ itself is never 
guaranteed. Fortunately, it is known that $M/M_{\rm bulge}$ of nearby AGNs 
is the same as that of local galaxies (Wandel 2002). 
It will be the first step to investigate the relation between the 
stellar mass and SMBH mass of LBGs which show the AGN activity.

{\acknowledgements I am grateful to S.Mineshige, Y. Ueda,
T.Miyaji, A.Inoue and N.Kawakatu for helpful 
comments and their encouragement. 
I thank M.Vestergaard for motivating me.
This work is supported 
in part by Research Fellowships of the 
Japan Society for the Promotion of Science for Young Scientists, 
05124.}

\clearpage

\clearpage

\clearpage

\clearpage

\clearpage

\end{document}